# Unraveling the optical contrast in $Sb_2Te$ and AgInSbTe phase-change materials

Shehzad Ahmed[1], Xudong Wang[1]*, Yuxing Zhou[1], Liang Sun[2], Riccardo Mazzarello[3,4]*, Wei Zhang[1,5]*

[1]Center for Alloy Innovation and Design (CAID), State Key Laboratory for Mechanical Behavior of Materials, Xi'an Jiaotong University, Xi'an 710049, China

[2]Key Laboratory of Materials Processing Engineering, College of Materials Science and Engineering, Xi'an Shiyou University, Xi'an 710065, China

[3]Department of Physics, Sapienza University of Rome, 00185 Rome, Italy

[4]Institute for Theoretical Solid-State Physics, JARA-FIT and JARA-HPC, RWTH Aachen University, 52056 Aachen, Germany.

[5]Pazhou Lab, Guangzhou, 510320, China

*Emails: xudong.wang@stu.xjtu.edu.cn, riccardo.mazzarello@uniroma1.it, wzhang0@mail.xjtu.edu.cn

**Abstract**

Chalcogenide phase-change materials (PCMs) show a significant contrast in optical reflectivity and electrical resistivity upon crystallization from the amorphous phase and are leading candidates for non-volatile photonic and electronic applications. In addition to the flagship $Ge_2Sb_2Te_5$ phase-change alloy, doped $Sb_2Te$ alloys, in particular AgInSbTe used in rewritable optical discs, have been widely investigated for decades, and nevertheless the theoretical insights on the optical properties of this important family of PCMs are scarce. Here, we carry out thorough ab initio simulations to gain an atomistic understanding of the optical properties of $Sb_2Te$ and AgInSbTe. We show that the large optical contrast between the amorphous and crystalline phase stems from the change in bond type in the parent compound $Sb_2Te$. Ag and In impurities serve mostly the purpose of stabilization of the amorphous phase, and have marginal impact on the large variation in the dielectric function upon the phase transitions.

# I. Introduction

Phase-change materials (PCMs) undergo fast and reversible transitions between crystalline and amorphous states exhibiting electrical and optical contrast, enabling a number of important applications.[1-8] The first commercial products of PCMs were rewritable optical media (CD, DVD and Blu-ray Disc), which exploited the reflectivity contrast and employed laser pulses to switch the material.[9-11] More recently, the large resistance difference between the amorphous and crystalline phases is being exploited in electronic devices such as the 3D Xpoint non-volatile memories,[12-15] where electrical pulses are used to induce the structural transitions. Such devices offer an attractive combination of properties including excellent scalability, stability, high speed, and non-volatility.[16-21] PCMs are also promising for applications in neuro-inspired computing devices, aiming to eliminate the von Neumann bottleneck of traditional computing.[22-30] Very recently, with the booming development of optoelectronics and photonics, the optical properties of PCMs have gained again strong interests. In addition to photonic non-volatile memories[31-36] and neuro-inspired computing,[37-40] various emerging techniques based on PCMs have been proposed and demonstrated, such as non-volatile optoelectronic displays, reconfigurable optical metamaterials, mid-infrared absorbers, thermal emitters and others.[41-49]

Important families of PCMs are type I—the pseudo-binary Ge-Sb-Te alloys $GeTe_{(1-x)}$–$(Sb_2Te_3)_x$, and type II—the doped $Sb_xTe$ (x≥2) binary compounds, such as Ag, In-doped $Sb_2Te$. The major difference between the two materials families is the nucleation rate. For instance, the nucleation time for the most widely studied $Ge_2Sb_2Te_5$ (GST) alloy is of the order of nanoseconds,[50] while that of AgInSbTe is of the order of microseconds.[51] For sub-micro-scale amorphous marks, the crystallization of GST proceeds via incubation of multiple nuclei and the subsequent grain growth, while AgInSbTe alloys typically crystallize via the amorphous-crystalline boundaries because of their high growth rate and low nucleation rate.[52] For the currently used memory cells of cell size of hundreds of nanometers, GST is widely used because the crystallization can be accomplished by simultaneous grain growth via multiple nucleation centers and interfacial growth from the amorphous-crystalline boundaries. The crystallization kinetics of PCMs has been thoroughly investigated by various experimental methods [50-60] and ab initio simulations. [60-68] It is generally agreed that the growth rate of AgInSbTe alloys could reach several meters per second at elevated temperatures, few times faster than that of GST alloys. Therefore, growth-type PCMs could become useful again in further miniaturized memory cells. A thorough review on the crystallization kinetics on PCMs is found in Ref. [2].

Regarding the optical contrast between the amorphous and crystalline phases for optical and photonic applications, a metavalent bonding – MVB framework [69-74] (previously known as resonant bonding [75]) has been developed. For instance, in the case of GST and GeTe, The amorphous phase shows mostly covalent bonding,[76-78] whereas the crystalline, rocksalt-like structure displays MVB, which is characterized by considerable electron delocalization.[69-74] In the metavalently bonded rocksalt phase, the *p* orbitals on neighboring atoms are aligned, which leads to the large matrix elements of the optical transitions and, thus, large optical absorption. On the contrary, the disordering and the misalignment of the p orbitals in the amorphous state decreases the optical matrix elements and the absorption, resulting in the optical contrast.[75, 79-83] Although the family of doped $Sb_xTe$ alloys has also been widely investigated for decades,[84-90] the optical properties of Sb$_2$Te and AgInSbTe are scarce. Here we perform ab initio simulations based on density functional theory (DFT) to determine the dielectric function of crystalline and amorphous Sb$_2$Te and AgInSbTe and elucidate their optical contrast. Similarly to GeSbTe compounds, we find that the contrast can be ascribed to subtle changes in bonding.

**II. Methods**
We carried out DFT-based ab initio molecular dynamics (AIMD) simulations using the second-generation Car-Parrinello molecular dynamics scheme[91] implemented in the CP2K package.[92] We employed the Perdew-Burke-Ernzerhof (PBE) functional[93] and the Goedecker pseudopotential.[94] The time step was set to 2 fs. The VASP package was also used to perform geometry relaxation and to compute the dielectric function. It was reported that hybrid functional calculations can satisfactorily reproduce the experimental optical contrast of typical phase-change materials, such as GeTe, GST and Sb$_2$Te$_3$.[81] Thus, we employed the projector augmented wave (PAW) method[95] in combination with both PBE and hybrid Heyd-Scuseria-Ernzerhof (HSE06) functionals.[96] For the crystalline structures, PBE simulations with van der Waals corrections[97] were also considered (Figure S1). An energy cut off of 550 eV for plane waves was used to achieve the desired accuracy. The structural relaxation was carried out with the conjugate gradient (CG) method. A grid of 15 × 15 × 3 k-points was used for c-Sb$_2$Te (whose primitive cell contains 9 atoms), whereas the $\Gamma$ point approximation was used for the 216 atoms supercells of amorphous Sb$_2$Te, and crystalline and amorphous Ag$_8$In$_6$Sb$_{144}$Te$_{58}$. Calculations with denser k-point meshes do not affect the conclusion on optical contrast between the amorphous and crystalline phases (Figure S2). The frequency-dependent dielectric matrix was calculated within the independent-particle approximation. Local field effects and many body effects were not considered, which was proven to be adequate to quantify the optical contrast between crystalline and amorphous GST.[81, 98] The dielectric function ε (ω) was determined in the energy interval of 0 to 7 eV. The imaginary part of ε (ω), termed as $\varepsilon_2(\omega)$, is given by

$$\varepsilon_2(\omega) = \frac{4\pi^2 e^2}{\Omega} \lim_{q \to 0} \frac{1}{q^2} \times \sum_{c,v,\mathbf{k}} 2w_k \delta(E_c - E_v - \omega)|\langle c|\mathbf{e} \cdot \mathbf{q}|v\rangle|^2, \quad (1)$$

where $\langle c|\mathbf{e} \cdot \mathbf{q}|v\rangle$ represents the joint optical transition from the states of the valence band (v) to the states of the conduction band (c), **e** is the direction of polarization of the photon, and **q** is the electron momentum operator.[99] The integration over the **k**'s is a sum over special k-points with corresponding weighting factors $w_\mathbf{k}$. In Formula (1), the matrix elements of the optical transitions are weighted by the Joint Density of States (JDOS) defined as:[100]

$$J(\omega) = 2 \sum_{v,c,\mathbf{k}} w_k \, \delta(E_c - E_v - \omega), \quad (2)$$

where $E_v$ and $E_c$ indicate the energies of the states in the valence and conduction bands, respectively.

**III. RESULTS AND DISCUSSIONS**
Crystalline (c-) Sb$_2$Te forms a rhombohedral structure (space group P$\bar{3}$m1) with alternate stacking of one Sb$_2$Te$_3$ quintuple-layer (QL) and two Sb bi-layers (BLs),[101, 102] as shown in Figure 1a. The DFT-optimized lattice parameters obtained by using the PBE functional are $a$ = 4.36 Å and $c$ = 17.77

Å. In the c-Sb$_2$Te, both Sb and Te atoms form octahedral bonding patterns. All the atoms display deviations from the perfect octahedral sites due to the pronounced Peierls distortion, yielding long-short Sb-Sb and Sb-Te bonding pairs. The Sb-Sb bond distance between the two Sb BLs is 3.43 Å, while the Sb-Te bond distance between the Sb BL and Sb$_2$Te$_3$ QL is 3.53 Å. The latter is smaller than the Te-Te bond distance between two QLs in rhombohedral Sb$_2$Te$_3$, 3.74 Å.[103] The structural gaps between QLs in rhombohedral Sb$_2$Te$_3$ are regarded as pseudo-van der Waals (vdW) gaps due to the presence of weak covalent interactions.[104] In rhombohedral Sb$_2$Te, the smaller bond distance between the Sb BLs and Sb$_2$Te$_3$ QL indicates stronger covalent interaction.

We carried out AIMD simulations to generate amorphous (a-) Sb$_2$Te models following the standard melt-quench protocol.[63] 144 Sb atoms and 72 Te atoms were placed in a cubic box, and were heated up to 3000 K to remove all possible crystalline order. The disordered model was then quenched down to and equilibrated at 1000 K over 30 ps, and was then quenched down to 300 K in 50 ps. After 30 ps equilibration at 300 K, the amorphous model was further quenched down to 0 K for electronic structure analyses and optical property calculations. The theoretical value of the supercell parameter is ~19.77 Å, corresponding to a density of 5.74 g/cm$^3$. A snapshot of amorphous Sb$_2$Te is shown in Figure 1b.

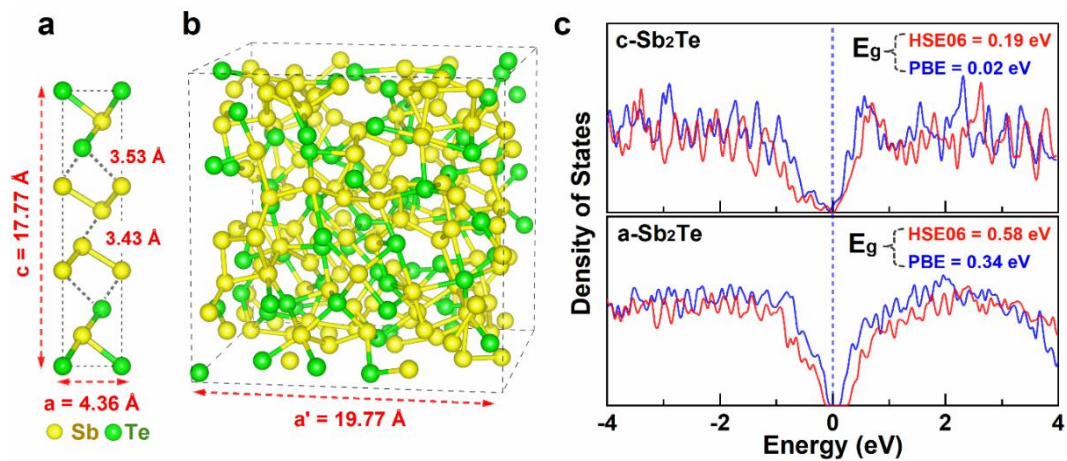

**Figure 1:** (a) The crystalline structure of rhombohedral Sb$_2$Te upon DFT-PBE relaxation. (b) The amorphous structure of Sb$_2$Te by melt-quenched AIMD simulation. Sb and Te atoms are rendered with yellow and green spheres. (c) Density of states (DOS) of crystalline and amorphous Sb$_2$Te calculated with both PBE and HSE06 functionals.

The crystalline and amorphous structures shown in Figure 1a-b were then used for the calculation of the electronic density of states (DOS). Besides the PBE functional, the hybrid HSE06 scheme was also employed to compute this quantity, since the latter typically gives more accurate band gap size, and more accurate optical properties. As shown in Figure 1c, a narrow band gap is present for both the crystalline and amorphous Sb$_2$Te. The gap size of c-Sb$_2$Te (0.19 eV, HSE06) is smaller than that of a-Sb$_2$Te (0.58 eV, HSE06).

Figure 2a shows the imaginary part of the dielectric functions $\varepsilon_2$ (i.e. the optical absorption) of crystalline and amorphous Sb$_2$Te. Significant differences are found between the two phases using either the PBE or the HSE06 functional. The $\varepsilon_2$ of the crystalline phase is much higher than that of the amorphous phase in the photon energy range from 0 eV to 3 eV, covering nearly the whole telecom wavelength and visible light spectrum. The peak value of the amorphous $\varepsilon_2$ curve is reduced by nearly three times and the peak position is shifted towards higher energies by ~0.5 eV with respect to those of the crystalline $\varepsilon_2$ curve. We note that inclusion of van der Waals corrections leads to variations in the $\varepsilon_2$ values in the telecom wavelength regions (C-band, ~0.8 eV), nevertheless the peak positions change very slightly. For the visible light region, the $\varepsilon_2$ curves hardly change (Figure S1). To explain this large contrast in $\varepsilon_2$, we compared the JDOS of the two phases in Figure 2b. The JDOS of c-$Sb_2Te$ is larger than that of a-$Sb_2Te$ in the energy range from 0 eV to 1 eV due to its smaller band gap, while the two curves approach each other at higher photon energies. Therefore, the JDOS alone cannot explain the large differences in $\varepsilon_2$ in the energy range from 1 eV to 3 eV.

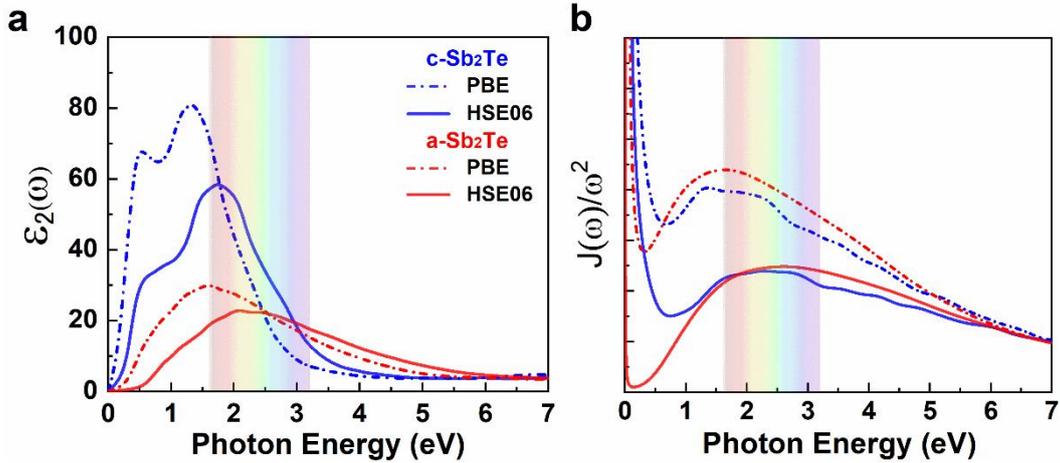

**Figure 2:** (a) Imaginary part of the dielectric function $\varepsilon_2$ and (b) $J(\omega)/\omega^2$ of the amorphous (red lines) and crystalline (blue lines) phases of $Sb_2Te$. Dashed and solid lines represent results calculated with PBE and HSE06 functionals, respectively. The rainbows in the figures highlight the visible light regions.

The observed optical contrast in $Sb_2Te$ is qualitatively similar to that found in GST. The latter was explained by the transition from MVB to covalent bonding upon amorphization of GST. In rocksalt GST, there are on average 3 $p$ electrons per lattice site (including 10% vacant sites), and the $p$ orbital alignment promotes electron delocalization, thus the large dielectric constant.[75] In contrast, amorphous GST is characterized by strong angular disorder between pairs of bonds,[81] which leads to $p$ orbital misalignment and covalent bonding, largely reducing the dielectric constant.[105] In fact, simulations showed that pronounced optical contrast is also obtained between two different crystalline structures of the parent PCM GeTe—the stable rhombohedral phase versus a hypothetical orthorhombic phase where strong $p$ orbital misalignment occurs.[80]

The angular distribution function (ADF) of a-$Sb_2Te$ is shown in Figure 3a. Octahedral bonds are typically found for both Sb and Te atoms with nearly 90° bond angles. On average Sb atoms have more neighbors (~4) than Te atoms (~3). The nearly aligned bond pairs result in an additional peak in ADF near 180°. Thorough structural analyses of a-$Sb_2Te$, including radial distribution function, coordination number, primitive rings statistics as well as cavity distribution, can be found in our earlier work.[106] Here, we focus on the bond alignment analysis. We first calculate the angular-limited three-body correlation (ALTBC)[107] to gain an overview about the bond correlation in a-Sb$_2$Te. We regard a pair of bonds with bond angle larger than 160° as a nearly aligned bond pair. The long/short bond pattern around all Sb atoms in a-Sb$_2$Te is mapped out in Figure 3b, showing a peak center around 2.9/3.5 Å. In c-Sb$_2$Te, three long/short bond correlations are present, namely, 3.04/3.20 Å for the Te-Sb-Te bond pair in the Sb$_2$Te$_3$ QL, 2.96/3.43 Å for the Sb-Sb-Sb bond pair in the Sb BLs, and 2.96/3.53 Å for the Te-Sb-Sb bond pair between the QL and BL.

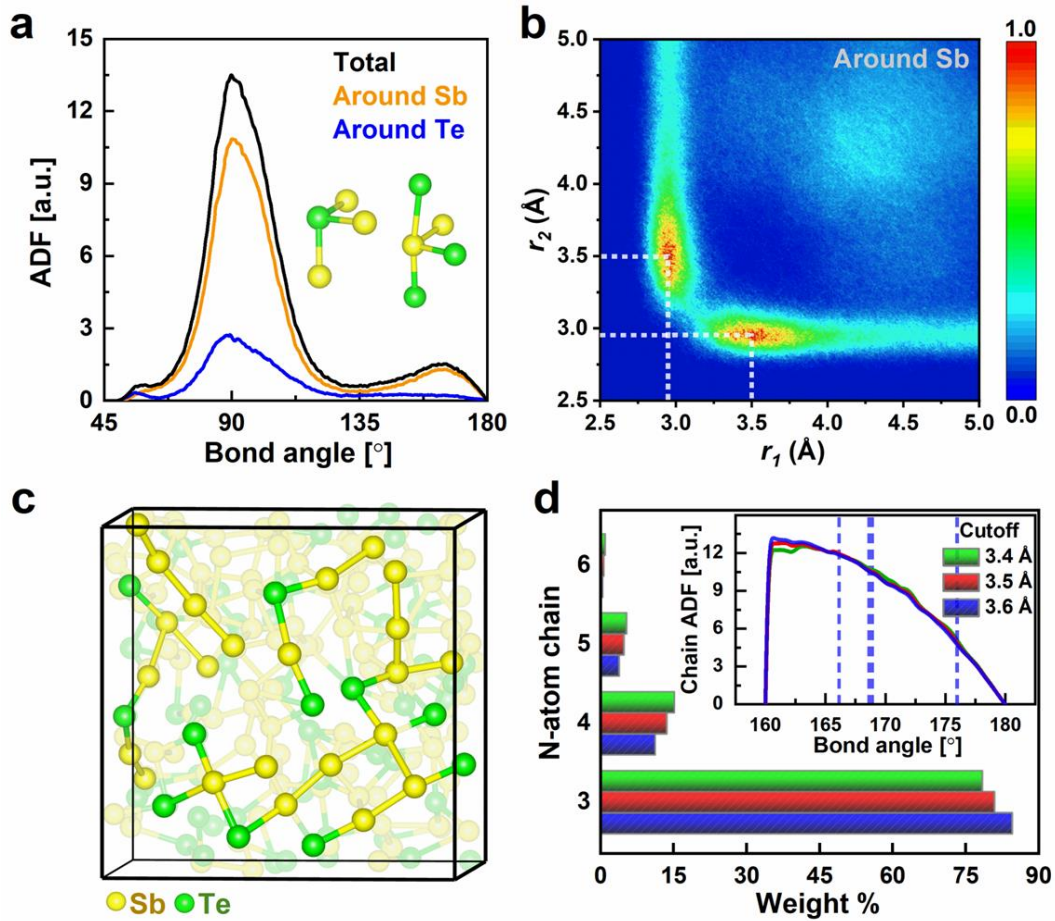

**Figure 3:** (a) The angle distribution function (ADF) and (b) the angular-limited three-body correlation (ALTBC) of a-Sb$_2$Te with bond angle larger than 160°. (c) A snapshot of a-Sb$_2$Te at 300 K highlights the long/short bond chains. (d) The corresponding analysis on chain length of nearly aligned bond pairs. The inset shows the ADF of the aligned bond chains. The dashed lines mark the ADF of the aligned bond chains in c-Sb$_2$Te. The structural data were collected at 300 K over 15 ps.

Next, we quantify the degree of angular disorder by analyzing the length of chains of aligned bond pairs. We collected and averaged the structural data over 15 ps at 300 K. The bond pairs with bond angles larger than 160° were considered for the chain analysis. Three distance cutoffs between 3.4 and 3.6 Å were tested, which showed qualitatively similar results (Figure 3b). Typical aligned bond chains are highlighted in Figure 3c. The detailed counting in Figure 3d shows that only ~18% of such chains can contain more than 3 atoms. The longest chain is identified to have 6 atoms. The angular distortion of these chains shows a broad distribution between 160° and 180° (Figure 3d inset). However, in c-Sb$_2$Te, despite slight angular distortions about 15° (as marked by blue dashed lines in Figure 3d inset), all the aligned bond chains are highly extended. Only lattice defects or grain boundaries can break these highly aligned bond chains. c-$Sb_2Te$ with both Sb BLs and $Sb_2Te_3$ QL shows on average 3 *p* electrons per lattice site locally, promoting electron delocalization and MVB. While in a-$Sb_2Te$, despite all atoms are octahedrally bonded, the strong angular disorder between these local motifs breaks the aligned bond chains, displaying conventional covalent bonding with poor electron delocalization.

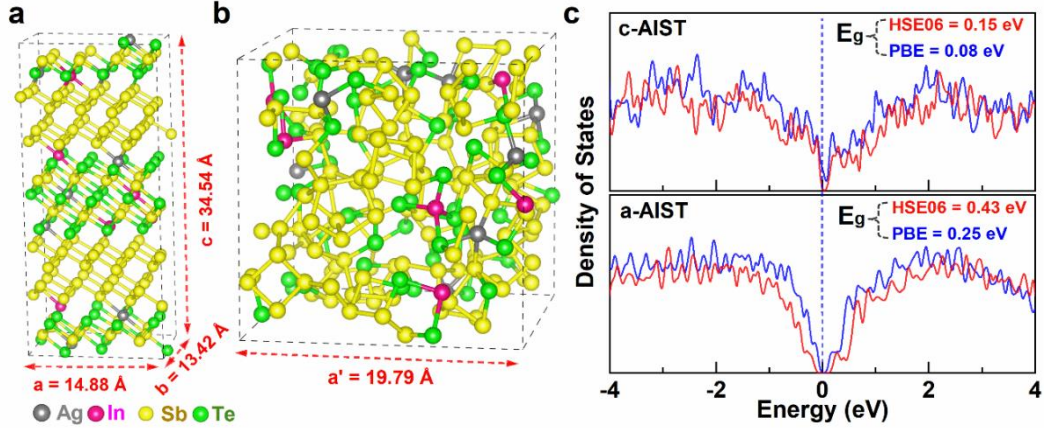

**Figure 4:** Atomic structures of (a) crystalline and (b) amorphous AIST. Ag, In, Sb and Te atoms are rendered with silver, pink, yellow and green spheres. (c) Electronic DOS of crystalline and amorphous AIST calculated with both PBE and HSE06 functionals.

Then we investigate the effects of Ag and In impurities on the optical properties. We chose a $Ag_8In_6Sb_{144}Te_{58}$ model (abbreviated as AIST in the following), which has a composition very close to the previous experimental and AIMD work presented in Refs. [52, 61] Since, for such composition, the ratio between Sb and Te atoms is about 2.48, Ag and In impurities are likely to occupy the positions of Te atoms in c-$Sb_2Te$. Figure 4a shows one possible c-AIST structure. The supercell was fully relaxed with the PBE functional. A cubic supercell with 216 atoms and the exact chemical composition was used to simulate the amorphous phase. Following the same melt-quench protocol, a-AIST was obtained, as shown in Figure 4b. These two atomic structures were then used for electronic structure calculations. The corresponding DOS plots are shown in Figure 4c. The presence of Ag and In impurities tends to reduce the size of band gap in both c-AIST (0.15 eV, HSE06) and a-AIST (0.43 eV, HSE06). The calculated band gap values of both phases are in good agreement with experimental values, which are 0.18 and 0.50 eV for c- and a-AIST, respectively.[108, 109] Projected DOS is shown in Figure S3 to further assess the contributions of the impurity orbitals to the electronic states.

Figure 5a shows the dielectric function $\varepsilon_2$ of crystalline and amorphous AIST. The peak height is reduced by ~24% (HSE06 data) for c-AIST with respect to c-$Sb_2Te$ (the trend is the same for PBE data). This is due to the fact that the presence of Ag and In impurities breaks the alignment of *p* orbitals in the crystalline phase, as Ag and In have, respectively, zero and one *p* electron in their valence shell. These impurities result in a weakening of MVB, and thereby a reduction of the dielectric function. Upon alloying the amorphous phase, the peak height of $\varepsilon_2$ is reduced by ~9% (HSE06 data) and the peak position is also shifted towards higher energies by 0.25 eV (HSE06 data). The change in $\varepsilon_2$ brought about by the impurities is smaller in the amorphous phase: this can be attributed to the fact that the aligned bond chains are anyway short in the amorphous phase. As regards the JDOS (Figure 5b), the difference between the crystalline and amorphous phase below 2 eV is more pronounced in AIST than in $Sb_2Te$, primarily because a very small band gap is found in c-AIST.

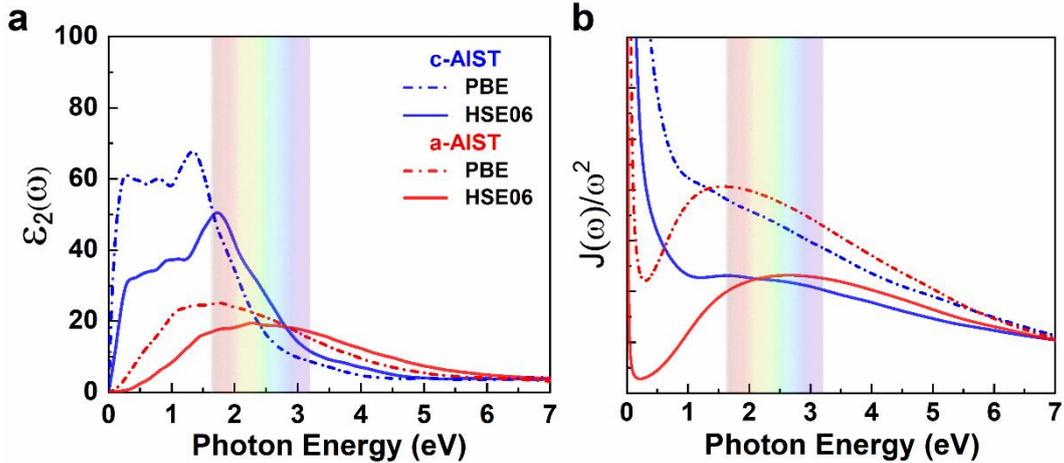

**Figure 5:** (a) Imaginary part of the dielectric function $\varepsilon_2$ and (b) $J(\omega)/\omega^2$ of the amorphous (red lines) and crystalline (blue lines) phases of AIST. Dashed and solid lines represent results calculated with both PBE and HSE06 functionals, respectively. The rainbows in the figures highlight the visible light regions.

At last, we note that the c-AIST structure considered here was chosen to resemble the ordered c-$Sb_2Te$ as much as possible to assess the impurity effects. In fact, simulations showed that, at elevated temperatures, AIST crystallizes on a nanosecond time scale in a A7 structure with random occupation of all four elements.[61, 85] It is expected that even $Sb_2Te$ crystallizes into a phase with strong substitutional Sb/Te disorder at high temperature. Upon long-term thermal annealing over hours, the Sb and Te atoms (and, possibly, the impurity atoms as well) can further rearrange to form a more stable structure with higher chemical order.[110] Nevertheless, the effects of Ag and In impurities should qualitatively be the same in the disordered crystalline phases, namely, breaking the aligned bond chains due to the shortage of *p* electrons.

## IV. CONCLUSION

In this work, we performed thorough DFT and AIMD simulations to investigate the optical contrast between amorphous and crystalline $Sb_2Te$ and AIST. We showed that the large imaginary part of the dielectric function in c-$Sb_2Te$ stems from highly extended bond chains of aligned *p* orbitals with 3 *p* electrons per lattice site on average, promoting electron delocalization and the formation of metavalent bonding. In contrast, strong angular disorder is present in a-$Sb_2Te$, breaking the aligned bond chains into smaller pieces and turning the bonding pattern into conventional covalent bonding. In addition, we assessed the impurity effects of common alloy elements, namely, Ag and In, which result in moderate weakening of absorption due to the breaking of the alignment of *p* orbitals in the crystalline phase. In spite of this, a significant optical contrast is also observed between crystalline and amorphous AIST. We conclude that the optical contrast in type II PCM originates mostly from its parent phase but not from the impurity atoms, due to their relatively low concentration. The impurities serve the purpose of enhancing the amorphous stability for practical applications.

In conclusion, this work elucidates a crucial property of $Sb_2Te$ and AIST relevant to existing applications. It also paves the way for further computational studies of alloyed $Sb_xTe$ alloys aiming at optimizing the material composition to tailor the optical properties for emerging applications, including neuro-inspired devices for scientific computing-oriented applications[27] and large-pixel reflective displays.[5] Doped $Sb_xTe$ alloys and other growth-type PCMs could offer several advantages as compared to nucleation-driven PCMs in these advanced applications, such as the lower stochasticity of the crystallization process due to the absence of nucleation, higher growth rate, and the capability of forming large crystalline grains.

**Acknowledgments**
W.Z. thanks the support of National Natural Science Foundation of China (61774123) and 111 Project 2.0 (BP2018008). S.A. acknowledges scholarship support from Chinese Scholarship Council. R.M. acknowledges funding from Deutsche Forschungsgemeinschaft (DFG) within SFB 917 ("Nanoswitches"). W.Z. acknowledges the support by the HPC platform of Xi'an Jiaotong University, and the International Joint Laboratory for Micro/Nano Manufacturing and Measurement Technologies of Xi'an Jiaotong University.

**V. SUPPORTING INFORMATION**

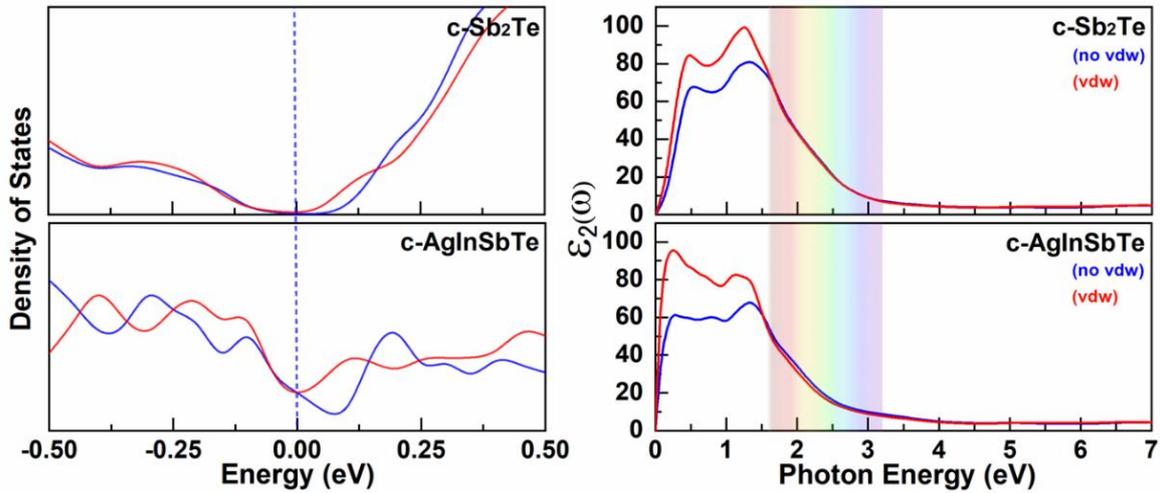

**Figure S1.** PBE calculations of DOS and $\varepsilon_2$ for both c-Sb$_2$Te and c-AIST with/without vdW corrections. In the visible light region (highlighted in the $\varepsilon_2$ figures), which is more important for practical applications, the $\varepsilon_2$ values with/without vdW corrections are nearly identical. Calculated with PBE functional.

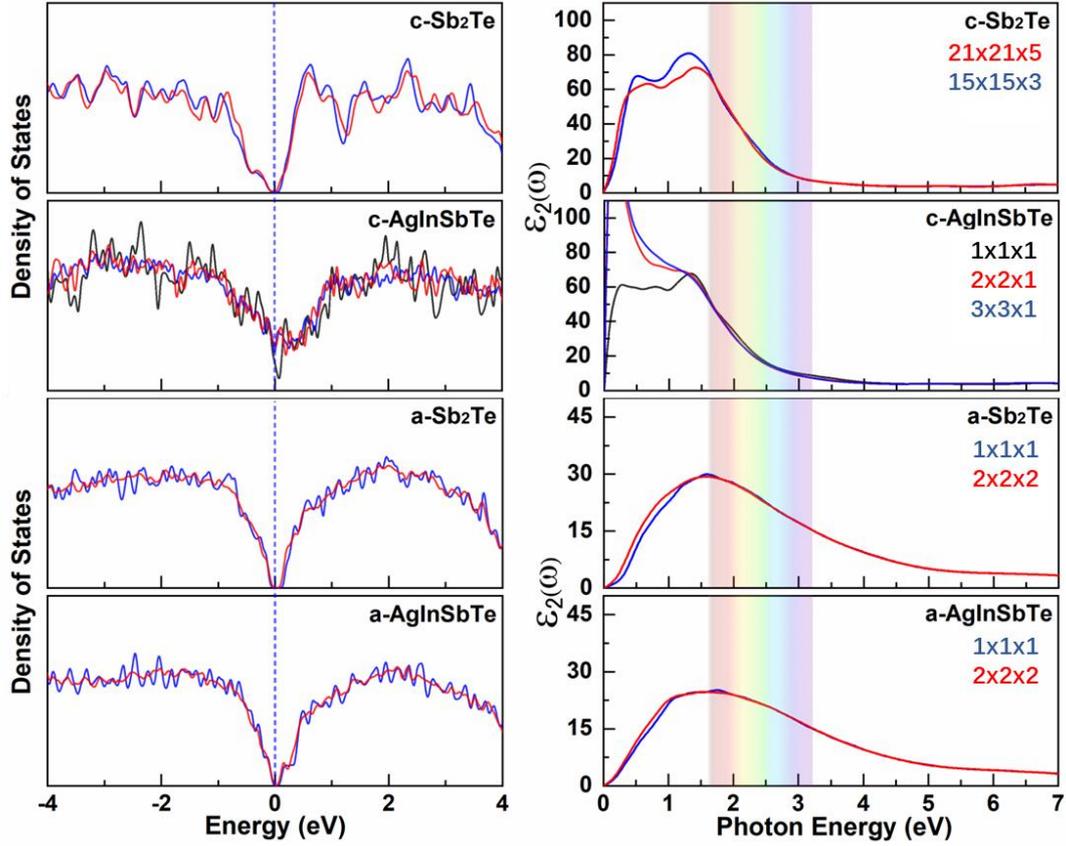

**Figure S2.** PBE calculations of convergence test for k points. $\varepsilon_2$ values are well converged in the visible light region for all the structures.

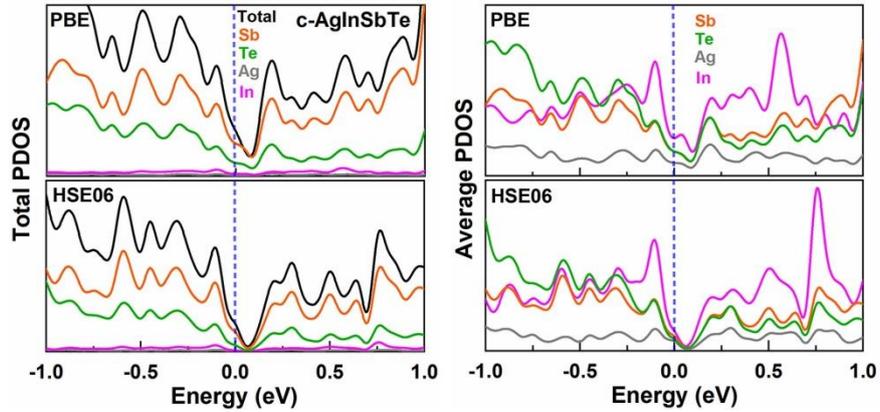

**Figure S3.** Projected DOS for each element in c-AIST. The average PDOS is obtained by dividing the respective PDOS over the number of atoms for each element.

13